\documentclass{article}
\usepackage{graphicx}
\usepackage{latexsym}
\usepackage{amsmath}
\usepackage{amsthm}
\usepackage{url,color}
\usepackage{bm}
\usepackage{caption}
\usepackage{subcaption} 
\usepackage{amssymb}

\newcommand{\vv}{{\bf v}}

\newcommand{\vu}{{\bf u}}

\newcommand{\vH}{{\bf H}}

\newcommand{\vR}{{\bf R}}
\newcommand{\bzero}{{\mathbf 0}}

\date{\today}

\newcommand{\be}{\begin{eqnarray}}
\newcommand{\ee}{\end{eqnarray}}

\newcommand{\R}{\mathbb{R}}
\newcommand{\half}{\frac{1}{2}}
\newcommand{\ep}{\varepsilon}

\newcommand{\om}{\Omega}
\newcommand{\Div}{{\rm div}\,}
\newcommand{\curl}{{\rm curl}\,}

\newcommand{\diam}{{\rm diam}\,}

\newcommand{\1}{{\bf 1}}
\newcommand{\tr}{{\rm tr}\,}

\newcommand{\mH}{{\mathcal H}}

\newcommand{\n}{{\bf n}}
\newcommand{\N}{{\bf N}}
\newcommand{\x}{{\bf x}}
\newcommand{\z}{{\bf z}}
\newcommand{\p}{{\bf p}}
\newcommand{\q}{{\bf q}}
\newcommand{\m}{{\bf m}}
\newcommand{\M}{{\bf M}}
\newcommand{\W}{{\bf W}}
\newcommand{\T}{{\bf T}}
\newcommand{\Q}{{\bf Q}}
\newcommand{\bphi}{\boldsymbol{\varphi}}

\newcommand{\az}{{\bf a}}

\newcommand{\e}{{\bf e}}

\newcommand{\y}{{\bf y}}
\newcommand{\A}{{\bf A}}
\newcommand{\D}{{\bf D}}
\newcommand{\Rz}{{\bf R}}
\newcommand{\nnu}{{\bm \nu}}
\newtheorem{thm}{Theorem}

\newtheorem{rem}{Remark}

\numberwithin{equation}{section}

\graphicspath{%
    {converted_graphics/}
    {C:/Users/John/Documents/Papers/}
    {/}
    {C:/Users/John/Pictures/}
}
\begin{document}
\title
{Mathematics and  liquid crystals}
\author{J. M. Ball\\ \\
Mathematical Institute,   University of Oxford,\\
Andrew Wiles Building,
Radcliffe Observatory Quarter,\\
Woodstock Road,
Oxford,
OX2 6GG,
 U.K. }

 \maketitle
 \markboth{ }{}
\begin{abstract}  \noindent 
A review is given  of some mathematical contributions, ideas and questions concerning liquid crystals.
\end{abstract}

\section{Introduction}
\setcounter{equation}{0}
Recent years have seen a remarkable increase in the number of mathematicians\footnote{In MathSciNet, the main mathematics reviewing journal, the number of papers mentioning liquid crystals rose from 4 for 1971-1975, to 223 for 2006-2010, and 407 for 2011-2015.} working on problems of liquid crystals. This reflects both the increased ability of modern mathematics to provide useful information on the problems of liquid crystals, and the interest and challenges of these problems for mathematicians. Indeed the  questions raised by liquid crystals are generating new mathematical techniques, which can be expected in turn to influence applications of mathematics to other parts of science.

Everyone who works on liquid crystals uses mathematics in one form or another, and professional mathematicians (for example, those who work in mathematics departments) don't have a prerogative on how mathematics should best be used. Nevertheless mathematicians can have a different way of viewing things that adds new perspectives, and it is one aim of this article to convey something of this.  The ultimate objective should be for mathematics to illuminate experiments and phenomena, to help predict and interpret experimental results, and to help suggest new experiments. Of course the hypotheses of mathematical results may often ignore relevant physical effects.  However by making simplifying assumptions it may be possible  to prove a rigorous theorem that both gives insight into what more complex models may predict, resolving  ambiguities arising from  other theoretical predictions based on approximations or numerical computation, and enabling placing of the problem within the existing mathematical body of knowledge.  

This article is intended less for mathematicians than for other researchers who may be interested in what mathematics can contribute to the field. Whereas I have attempted to give clear statements of results, at some points these are more informal than would be the case for a mathematical audience.
\section{The Oseen-Frank model and the description of defects}
Consider a nematic liquid crystal at rest at constant temperature in the absence of applied electromagnetic fields. (Although most of the applications of liquid crystals to technology depend  on their interaction with electromagnetic fields, for simplicity such interactions are  not discussed in this paper, since many of the key analytic difficulties already arise in the case when such fields are absent.) 

Suppose that the liquid crystal completely fills a container $\Omega\subset\R^3$, which we assume to be a bounded open set with boundary $\partial\Omega$. The configuration of the liquid crystal at the point $\x$ is described by the {\it director} $\n(\x)$, a unit vector  giving the mean orientation  at  $\x$ of the rod-like molecules comprising the liquid crystal. Thus $\n(\x)\in S^2$, where $S^2$ denotes the the unit sphere of $\R^3$.

\subsection {The Oseen-Frank energy}
In the Oseen-Frank model the total free energy is given by 
\be\label{OF}I(\n)=\int_\Omega W(\n,\nabla\n)\,d\x,
\ee
where
\begin{eqnarray} \label{OFE}  &&\hspace{-.3in}W(\n,\nabla\n) = K_{1}(\textrm{div}\,
\n)^{2} + K_{2}(\n\cdot \textrm{curl}\,\n)^{2}  +
K_{3}|\n \wedge \textrm{curl}\,\n|^{2}\\&&\hspace{2in}  +
(K_{2}+K_{4})(\textrm{tr}(\nabla \n)^{2} - (\textrm{div}\,
\n)^{2}).\nonumber
\end{eqnarray}
Here $K_1,\ldots, K_4$ are the {\it Frank constants}, which are usually assumed to satisfy the strict form of the  Ericksen inequalities \cite{ericksen1966}
\be 
\label{ericksen}
K_1>0,\, K_2>0,\, K_3>0,\, K_2>|K_4|, \,2K_1>K_2+K_4,
\ee
which are necessary and sufficient that
\be 
\label{lowerbound}
W(\n,\nabla\n)\geq c_0|\nabla\n|^2
\ee for all $\n$ and some constant $c_0>0$.
The first three terms in \eqref{OFE} describe respectively {\it splay, twist} and {\it bend} of the director field. The fourth {\it saddle-splay} term is a {\it null Lagrangian}, its integral
$$(K_2+K_4)\int_\Omega (\textrm{tr}(\nabla \n)^{2} - (\textrm{div}\,
\n)^{2})\,d\x$$
depending only on the values of $\n$ on the boundary $\partial\Omega$. Thus if $\n|_{\partial\Omega}$ is prescribed, such as for homeotropic or planar boundary conditions, the term can be ignored. However this is  not the case if $\n|_{\partial\Omega}$ is only partially prescribed, such as for the planar degenerate boundary condition $\n(\x)\cdot\nnu(\x)=0$ on part of the boundary having normal $\nnu(\x)$ (see, for example, \cite{koning2014}), or for weak anchoring boundary conditions.

An important identity is
\be 
\label{identity}|\nabla\n|^2=(\Div\n)^2+(\n\cdot\curl\n)^2+|\n\wedge\curl\n|^2+(\tr (\nabla\n)^2-(\Div\n)^2).
\ee
So if $K_1=K_2=K_3, K_4=0$ (the \it one-constant approximation\rm)  then 
\be 
\label{oneconstant}I(\n)=K_1\int_\Omega |\nabla \n|^2d\x,
\ee
which is the energy functional for {\it harmonic maps}.

\subsection {Orientability}
\begin{figure}[h] 
  \centering
  \includegraphics[width=4.5in,height=4in,keepaspectratio]{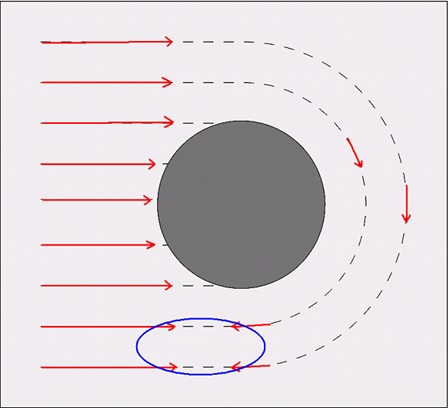}
   \caption{A smooth line field in the exterior of a cylinder that cannot be oriented to make it a continuous vector field.  The line field is parallel to the curves shown, with zero component in the perpendicular direction, so that the problem is two-dimensional. Indicated  is an attempt to orient the line field,  leading to a conflict in the  ellipse  shown.}
  \label{cylinder}
\end{figure}

The Oseen-Frank theory regards $\n$ as a vector field. However, due to the statistical head-to-tail symmetry of the constituent molecules, from a physical point of view $\n(\x)$ is indistinguishable from $-\n(\x)$. Hence $\n(\x)$ is better thought of as a {\it line field}, or equivalently
as a map from $\Omega$ to the set of
all lines through the origin.  A typical such line  
with direction $\pm\n$ can be represented by the matrix
$\n\otimes \n$ having components $n_in_j$.  The set of such lines
forms the
{\it real projective plane} $\R P^2$. 
Continuous (even smooth) line fields need not be orientable, so that it is 
impossible to assign a direction
that turns them into a continuous vector field (see Fig. \ref{cylinder}).
We return later to the implications of
this for the Oseen-Frank theory. For a detailed discussion see \cite{j57,j61}.
\subsection{Energy minimization and the Euler-Lagrange equation}
The fundamental {\it energy minimization problem} is to 
 find $\n$ that minimizes $$I(\n)=\int_\Omega W(\n,\nabla\n)\,d\x$$ subject to the unit vector constraint $|\n|=1$ and suitable 
boundary conditions, for example $\n|_{\partial\Omega}=\bar\n$,
where $\bar\n$ is  given. This is a problem of the {\it calculus of variations}.

If $\n$ is a minimizer and  $\m:\Omega\to\R^3$ is any smooth
mapping vanishing in a neighbourhood of $\partial\Omega$ then, for sufficiently small $|\varepsilon|$, 
\be\label{outer}\n_\varepsilon(\x)=\frac{\n(\x)+\varepsilon\m(\x)}
{|\n(\x)+\varepsilon\m(\x)|}\ee satisfies $|\n_\varepsilon(\x)|=1$ and $\n_\varepsilon|_{\partial\Omega}=\bar\n$.
Since $I(\n_\ep)$ is minimized at $\ep=0$  we have that $\frac{d}{d\varepsilon}I(\n_\varepsilon)|_{\varepsilon=0}=0$, provided this derivative exists. 
Noting that $$\frac{d}{d\varepsilon}\n_\varepsilon(\x)|_{\varepsilon=0}=({\bf 1}-\n(\x)\otimes \n(\x))\m(\x),$$
we obtain the {\it weak form} of the Euler-Lagrange equation, that for all such $\m$
\begin{eqnarray*}\int_\Omega\left(\frac{\partial W}{\partial\nabla\n}\cdot \nabla(({\bf 1}-
\n(\x)\otimes \n(\x))\m(\x))\right.&&\hspace{1.6in}{(WEL)}\\ &&\left.  \hspace{-.5in}+\frac{\partial W}{\partial\n}\cdot({\bf 1}-
\n(\x)\otimes \n(\x))\m(\x)\right)\,d\x=0,
\end{eqnarray*}
or, in components,
$$\int_\om\left(\frac{\partial W}{\partial n_{i,j}}((\delta_{ik}-n_in_k)m_k)_{,j}+\frac{\partial W}{\partial n_i}(\delta_{ik}-n_in_k)m_k\right)\,d\x=0,$$
where repeated indices are summed from 1 to 3. Hence, integrating by parts and using the arbitrariness of $\m$,
we formally obtain the 
Euler-Lagrange equation
$$\hspace{.8in}({\bf 1}-\n\otimes\n)\left({\rm div}\,\frac{\partial W}{\partial \nabla \n}-\frac{\partial W}{\partial \n}\right)=\bzero,\hspace{1in}(EL)$$
a system of second order nonlinear PDE to be solved subject to the pointwise constraint $|\n|=1$.

This can be written in the equivalent form
\be \label{lagmult}{\rm div}\,\frac{\partial W}{\partial \nabla \n}-\frac{\partial W}{\partial \n}=\lambda(\x)\n,\ee
where $\lambda(\x)$ is a Lagrange multiplier, or in components
$$\frac{\partial}{\partial x_j}\left(\frac{\partial W}{\partial n_{i,j}}\right)-\frac{\partial W}{\partial n_i}=\lambda(\x)n_i.$$

\noindent In the one-constant approximation $(EL)$ becomes the harmonic map equation
\be\label{harm}\Delta \n+|\nabla\n|^2\n=\bzero.\ee
How can we solve these equations? Are there any exact solutions? 
\subsection{Universal solutions}
\label{universal}
The question of what  $\n(\x)$ can be  solutions of $(EL)$ (smooth in some region of $\R^3$)
for {\it all} $K_1, K_2, K_3, K_4$, so called {\it universal solutions},
was addressed by Marris \cite{marris1978,marris1979}, following  Ericksen \cite{ericksen1967}. Marris showed that these consist of \\  

\noindent (i) constant vector fields, or those orthogonal to families of 
concentric spheres or cylinders,\\
(ii) pure twists, such as
\be \label{puretwist}\n(\x)=(\cos \mu x_3, \sin \mu x_3,0),\ee
(iii) planar fields that form concentric or coaxial circles.\\

An example from family (i) is the {\it hedgehog}
\be\label{hedgehog}\hat\n(\x)=\frac{\x}{|\x|}.\ee
which represents a point defect. Of course $\hat\n$ is not even continuous at $0$, but for $\x\neq 0$ it is smooth and we have
$$\nabla\hat\n(\x)= \frac{1}{|\x|}\left({\bf 1}-\frac{\x}{|\x|}\otimes\frac{\x}{|\x|}\right), \; |\nabla\hat\n(\x)|^2=\frac{2}{|\x|^2},$$ so that formally  
calculating its energy over the ball $B=\{\x\in\R^3:|\x|<1\}$, noting that $W(\n,\nabla\n)\leq C|\nabla\n|^2$ for some constant $C>0$,  we find that 
$$\int_BW(\hat\n,\nabla\hat\n)\,d\x\leq C\int_B|\nabla\hat\n|^2d\x =4\pi C\int_0^1r^2\cdot\frac{2}{r^2}\,dr<\infty.$$
More precisely, $\hat\n$ belongs to the {\it Sobolev space}
\be\label{H1}H^1(\Omega;S^2)=\{\n:\Omega\to S^2: \int_\Omega |\nabla\n|^2\,d\x<\infty\}.\ee
For a precise definition of  Sobolev spaces the reader is referred to standard texts on partial differential equations, such as \cite{evanspde}. See also the discussion in \cite{j57}. The most important point in giving a precise definition is that  $\nabla \n$ has to be defined in a suitable weak sense. For the purposes of this article one can informally think of $H^1(\Omega;S^2)$ as the space of finite-energy configurations for the Oseen-Frank theory; it is an example of a
{\it function space}, that specifies allowed singularities in a mathematical model.

\subsection{Existence of solutions and their regularity}
A routine use of the so-called `direct method' of the calculus of variations gives:
\begin{thm}
\label{existence} Let $K_1>0, K_2>0, K_3>0$, and let  $\bar\n\in H^1(\om;S^2)$.  Then there exists
 $\n^*$ that minimizes $I(\n)$ over all
$\n\in H^1(\Omega;S^2)$ with $\n|_{\partial\Omega}=\bar\n$, and any such minimizer 
$\n^*$ satisfies $(WEL)$.
\end{thm}
\begin{rem}\rm Because $\n$ is specified on the whole of $\partial\om$ and because the saddle-splay term is a null Lagrangian, we do not need any conditions on $K_4$ to get the existence of a minimizer. However, for other boundary conditions such as planar degenerate we would need to assume the Ericksen inequalities \eqref{ericksen}.
\end{rem}
\noindent Theorem \ref{existence} just tells us that there is {\it some} energy-minimizing director configuration $\n^*$, that is $I(\n)\geq I(\n^*)$ for all $\n$ with $\n|_{\partial\Omega}=\bar\n$, and that any minimizer satisfies $(WEL)$, and not what minimizers look like. Nevertheless the conclusion of the theorem is not obvious. Nor is it obvious that just because the problem posed has a sound physical basis the existence of a minimizer is necessarily assured, examples to the contrary arising, for example, in models of martensitic phase transformations \cite{j32}, where nonattainment of minimum energy leads to an understanding of the appearance of very fine microstructures. 

In the one-constant approximation \eqref{oneconstant} with $K_1>0$ there are deep results asserting more.
\begin{thm}[Schoen \& Uhlenbeck \cite{schoenuhlenbeck}, Brezis, Coron \& Lieb \cite{breziscoronlieb}]\label{finitenumber}
In the one-constant approximation with $K_1>0$ any minimizer 
$\n^*$ is smooth in $\om$ except for a finite number of point defects
located at points $\x(i)\in\Omega$, and
$$\n^*(\x)\sim \pm {\bf R}(i)\frac{\x-\x(i)}{|\x-\x(i)|}\mbox{ as }\x\to \x(i),$$
for some ${\bf R}(i)\in SO(3)$.
\end{thm}
\begin{thm}[Brezis, Coron \& Lieb \cite{breziscoronlieb}] \label{hedgehogmin}In the one-constant approximation the hedgehog $\hat\n$ minimizes $I$ subject to its own boundary conditions.
\end{thm}
\noindent There is an elegant alternative proof of Theorem \ref{hedgehogmin}   due to Lin \cite{lin1987}. The discovery of Theorems \ref{finitenumber} and \ref{hedgehogmin} was motivated in particular by experiments concerning point defects and their annihilation \cite{williamspieranskicladis1972}, and by corresponding analytical and numerical studies \cite{cohenetal1987,hardtkinderlehrerlin1986,hardtkinderlehrerlin1988,hardtkinderlehrerluskin1988}.

What about other  solutions to $(WEL)$ (these are often called {\it weak solutions} to $(EL)$)? Unfortunately there seem to be far too many.  In the one-constant approximation Rivi\`ere \cite{riviere1995}
 showed that for a domain $\Omega$ with smooth boundary, with $\bar\n$ smooth on $\partial\om$ and not constant, there are infinitely
many weak solutions, and that they can be discontinuous everywhere. The idea behind the construction is to insert dipoles (pairs of  point defects
  with opposite topological charge) into a non constant smooth map. This is reminiscent of other physical systems of  nonlinear partial differential equations where
there are too many solutions  and  no satisfactory way of selecting the 
physical one, such as exhibited by Scheffer \cite{Scheffer1993}, Shnirelman \cite{shnirelman1997} and De Lellis 
\& Sz\'ekelyhidi \cite{delellisszekelyhidi2009} for the Euler equations
of inviscid flow.

The derivation of $(WEL)$ used the {\it outer variation} \eqref{outer}. However, one can also take the {\it inner variation}
\be 
\label{inner}
\n_\ep(\x)=\n(\z_\ep(\x)),
\ee 
where   $\bphi:\om\to\R^3$ is a smooth function vanishing in a neighbourhood of $\partial\om$, and $\z_\ep(\x)$ is implicitly defined by
\be 
\label{var}\z_\ep(\x)+\ep\bphi(\z_\ep(\x))=\x.\ee
This variation rearranges the values of $\n$ in $\om$ and preserves the unit vector constraint.
For sufficiently small $|\ep|$   \eqref{var} has a smooth solution $\z_\ep:\om\to\om$ with $\z_\ep(\x)=\x$ near $\partial\om$. Changing variables from $\x$ to $\z=\z_\ep$ we see that 
\be 
\label{newI}
I(\n_\ep)=\int_\om W(\n(\z),\nabla\n(\z)(\1+\ep\nabla\bphi(\z))^{-1})\det(\1+\ep\nabla\bphi(\z))\,d\z.
\ee Since $I(\n_\ep)$ is minimized at $\ep=0$ we have that $\frac{d}{d\ep}I(\n_\ep)|_{\ep=0}=0$, giving  
\be\label{innervar}\int_\om\left(W\1-(\nabla\n)^T\frac{\partial W}{\partial \nabla\n}\right)\cdot\nabla{\bphi}\,\,d\x=0\ee
for all such $\bphi$. This is a weak form of the equation
\be\label{em}
\Div\left(W\1-(\nabla\n)^T\frac{\partial W}{\partial \nabla\n}\right)=\bzero.
\ee
If $\n$ is a smooth solution of $(EL)$ then it is easily verified that $\n$ also satisfies \eqref{em}. However solutions of $(WEL)$ do not necessarily satisfy \eqref{innervar}. Solutions of $(WEL)$ that also satisfy $\eqref{innervar}$ are called {\it stationary}. In the one-constant approximation it was proved by Evans \cite{evans1991} that any stationary solution is partially regular, that is such a solution is smooth outside a closed subset $E$ of $\om$ having zero one-dimensional Hausdorff measure. Thus $E$ cannot, for example, contain a line segment or arc. In view of the result of Rivi\`ere mentioned above, this means that stationary solutions of $(WEL)$ are more regular than general solutions.  

\subsection{Dynamics and weak equilibrium solutions}
\label{dynamics}
It is interesting to examine the dynamical significance of the different notions of weak solution\footnote{I am grateful for discussions with Fanghua Lin and Epifanio Virga conerning the material in this section.}. The relevant dynamical equations are the Ericksen-Leslie equations (see \cite{ericksen62, leslie1966,leslie1968} and for a modern treatment \cite[Chapter 3]{SonnetVirga2012}). For an incompressible fluid in the absence of body forces and couples, these equations for the velocity $\vv(\x,t)$ and director $\n(\x,t)$ take the form
\be 
\label{ericksenleslie}
 \rho\left(\frac{\partial\vv}{\partial t}+(\vv\cdot\nabla)\vv\right) -\Div\T&=&\bzero,\\  \Div\vv&=&0,\label{inc}\\
(\1-\n\otimes\n)\left(\frac{\partial R}{\partial\mathring \n}+\frac{\partial W}{\partial\n}-\Div \frac{\partial W}{\partial\nabla\n}\right)&=&\bzero\label{ericksenleslie1},
\ee
where $\rho$ is the constant density,
 $\T$ is the Cauchy stress tensor, $R=R(\n;\D,\mathring \n)$ is the dissipation potential, $\mathring \n=\left(\frac{\partial\n}{\partial t}+(\vv\cdot\nabla)\n\right)-\W\n$, $\D=\frac{1}{2}[\nabla\vv+(\nabla\vv)^T]$ and $\W=\frac{1}{2}[\nabla\vv-(\nabla\vv)^T]$. For the purposes of this discussion all we need to know is that when $\vv=0$ and $\n=\n(\x)$,
\be 
\label{Tequil}
\T=-p\1-(\nabla\n)^T\frac{\partial W}{\partial\nabla\n},\;\;\frac{\partial R}{\partial\mathring \n}=\bzero,
\ee
where $p$ is the pressure (a Lagrange multiplier corresponding to the incompressibility constraint \eqref{inc}). Note that it is indeed the dynamical equations for the flow of an {\it incompressible} liquid crystal that are appropriate here. Were we to consider  compressible flow then the corresponding free-energy density $W$ would depend not only on $\n, \nabla\n$ but also on the density $\rho$ and perhaps $\nabla\rho$ (see \cite[Chapter 5]{SonnetVirga2012}), leading to a different variational problem for a functional $I(\rho,\n)$. 

The requirement that $\vv=0$, $\n=\n(\x)$, $p=p(\x)$ (where we assume that $p$ is integrable over $\om$) is a weak solution of \eqref{ericksenleslie}-\eqref{ericksenleslie1} is then that $(WEL)$ holds and that 
\be\label{equil}\int_\om\left(p\1+(\nabla\n)^T\frac{\partial W}{\partial \nabla\n}\right)\cdot\nabla{\bphi}\,d\x=0,
\ee
for all smooth $\bphi$ vanishing in a neighbourhood of $\partial\om$. If $\n$ is a stationary solution of $(WEL)$ then \eqref{equil} holds with $p=p_0-W$, where $p_0$ is constant. However, it is not clear whether or not a solution $\n$ of $(WEL)$ satisfying \eqref{equil} is stationary. This would follow if we knew that $\n$ were partially regular, but this is also unclear. 

It is interesting (see Ericksen \cite{ericksen1968twist}) that \eqref{ericksenleslie} can admit solutions with $\vv=\bzero$ but $\n=\n(\x,t), p=p(\x,t)$, for which similar considerations apply. 

\subsection{Line defects}
Another solution of type (i) is the two-dimensional hedgehog given by 
\be 
\label{2Dhedgehog}\tilde \n(\x)=(\frac{x_1}{r},\frac{x_2}{r},0),\;\;r=\sqrt{x_1^2+x_2^2}.
\ee 
This has the {\it line defect} $L=\{(0,0,z): z\in\R\}$. However,  for any bounded domain $\Omega$ containing part of $L$,   we have that some cylinder $$\{\x=(x_1,x_2,z): a<z<a+L, \sqrt{x_1^2+x_2^2}<r_0\}$$ is contained in $\Omega$, so that using \eqref{lowerbound} 
\be 
\label{infinite}
I(\n)\geq 2\pi c_0L\int_0^{r_0}r\cdot\frac{1}{r^2}\,dr=\infty.
\ee
Thus the two-dimensional hedgehog has infinite energy, and $\tilde\n\not\in H^1(\Omega;S^2)$.

\begin{figure}[h] 
\begin{subfigure}{0.2\textwidth}
\includegraphics[scale=0.58]{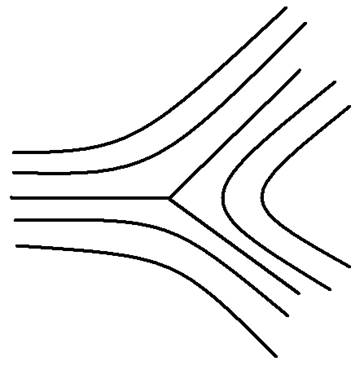}\caption{}
\end{subfigure} \hspace{.4in} 
\begin{subfigure}{0.2\textwidth}
\includegraphics[scale=0.58]{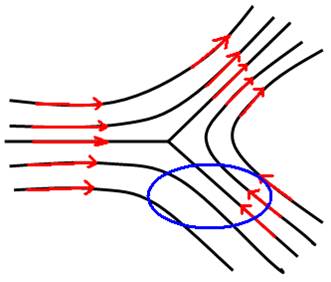}\caption{}
\end{subfigure} \hspace{.6in}
\begin{subfigure}{0.2\textwidth}
\includegraphics[scale=0.58]{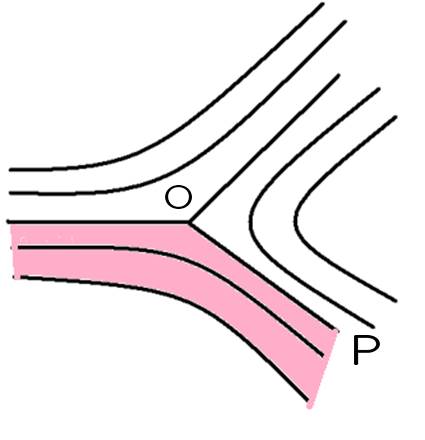}\vspace{-.3in}\caption{}
\end{subfigure}
\caption{An index $\frac{1}{2}$ defect. }
\label{indexhalf}
\end{figure}

 Other line defects  are given by {\it index $\frac{1}{2}$ defects}, such as that illustrated in Fig. \ref{indexhalf}(a). Here the director lies in the $(x_1,x_2)-$plane and is parallel to the curves shown, with zero component in the $x_3$ direction. Such configurations cannot be described by the Oseen-Frank theory for two reasons. First, they are not orientable (see Fig. \ref{indexhalf}(b), where an attempt to orient it is illustrated, giving a conflict in the elliptical cylinder shown) so that  they cannot be represented by a vector field $\n$ that is continuous outside the defect. Second, even if we restrict attention to one sector in which the line-field is orientable (such as the shaded region in  Fig. \ref{indexhalf}(c)) the corresponding energy is infinite. 

The problem that $\tilde\n$ and the index $\frac{1}{2}$ singularities have infinite energy could potentially be fixed by modifying the growth of $W(\n,\nabla\n)$ for large $|\nabla\n|$ to be {\it subquadratic}, i.e. $W(\n,\nabla\n)\leq C(1+|\nabla\n|^p)$ for some $C>0, 1\leq p<2$. After all, the quadratic dependence of $W$ in $\nabla\n$ suggests a theory designed to apply for {\it small} values of $|\nabla\n|$, whereas near defects $|\nabla\n|$ is very large. As described in \cite{j67} it is easy to modify $W$ to have subquadratic growth without affecting the Frank constants or other desirable properties. We return in Section \ref{lavrentiev} to the question of how the lack of orientability of the index $\frac{1}{2}$ defect might be handled.

\section{The Lavrentiev phenomenon and function\\ spaces}
\label{lavrentiev}
 The {\it Lavrentiev phenomenon}, discovered in 1926 by Lavrentiev \cite{lavrentiev26}, is of profound importance for mechanics and physics. It occurs even in harmless looking problems of the one-dimensional calculus of variations\footnote{for example, the problem of minimizing the integral $I(u)=\int_{-1}^1[(x^4-u^6)^2u_x^{28}+\varepsilon u_x^2]\,dx$ subject to the boundary conditions $u(-1)=-1, u(1)=1$, where $\ep>0$ is sufficiently small. For a detailed discussion, together with other historical references, see \cite{j28}.}. It can be expressed as the statement:\\

\fbox{\begin{minipage}{29em}
Minimizers of the same energy in different function spaces can be different,
and give different values for the minimum energy.
\end{minipage}}\\

\subsection{An example from solid mechanics}
\begin{figure}[h]   
\begin{subfigure}{0.15\textwidth}  \begin{center}\vspace{.2in}
\includegraphics[scale=0.58]{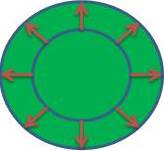} \caption{ }\vspace{.2in}\end{center}
\end{subfigure} \hspace{.5in} 
\begin{subfigure}{0.15\textwidth}
\includegraphics[scale=0.58]{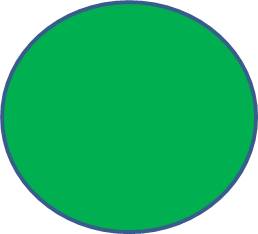}\hspace{1in}\caption{}
\end{subfigure} \hspace{.43in}
\begin{subfigure}{0.15\textwidth}
\includegraphics[scale=0.58]{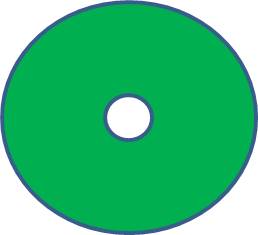} \caption{}
\end{subfigure}
\hspace{.4in}
\begin{subfigure}{0.15\textwidth}
\includegraphics[scale=0.58]{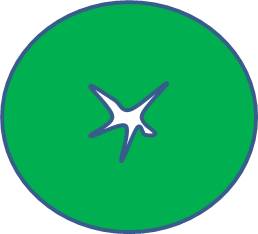}\hspace{.5in}\caption{}
\end{subfigure}
\caption{Solutions to the radial expansion problem for an elastic ball in different function spaces. (a) shows the boundary conditions, (b) the minimizer in the
class of smooth maps, (c) the minimizer among radial maps in $H^1(\om;\R^3)$, with a cavity at the origin, (d) a possible minimizer in the space $SBV(\om;\R^3)$.  }
\label{rubber}
\end{figure}

 For physical systems the Lavrentiev phenomenon  has the uncomfortable implication that {\it the function space is part of the model}.  An interesting illustration of this in a physical problem  comes from nonlinear elasticity theory. Consider a nonlinear elastic body (composed of rubber, say) occupying in a reference configuration the unit ball $B=\{\x\in\R^3:|\x|<1\}$. A typical deformation is described by a map $\y:B\to\R^3$, where $\y(\x)$ denotes the deformed position of the material point $\x\in B$.  We look for a minimizer $\y$ of the total elastic free energy
\be 
\label{elastic}
I(\y)=\int_B W(\nabla\y(\x))\,d\x,
\ee
subject to the boundary condition 
\be
\label{ballbc}
\y(\x)=\lambda\x, \;\;\;\x\in\partial B,
\ee
where $\lambda>0$,
corresponding to a uniform radial expansion at the boundary (see Fig. \ref{rubber}(a)). We take as an example the compressible neo-Hookean material having free-energy density
\be 
\label{neohookean}
W(\A)=\mu|\A|^2+h(\det\A),
\ee
where $\mu>0$ and $h:(0,\infty)\to [0,\infty)$ is a convex function satisfying 
\be 
\label{growthelastic}\lim_{\delta\to 0+}h(\delta)=\lim_{\delta\to\infty}h(\delta)/\delta=\infty,
\ee
 which is a standard model (though not the best) for rubber. 
Among smooth mappings $\y$ the unique minimizer of $I(\y)$ subject to \eqref{ballbc} is given (see Fig. \ref{rubber}(b)) by the uniform dilatation $\y(\x)=\lambda\x$ (this follows from the fact that, due to the convexity of $h$, $W$ is {\it polyconvex}, hence {\it quasiconvex}; for the explanation of these terms and more details see \cite{p31}). However, if $\lambda$ is sufficiently large, the minimizer in the Sobolev space $H^1(\om;\R^3)=\{\y:\om\to\R^3:\int_\om(|\y|^2+|\nabla\y|^2)\,d\x<\infty\}$ is not given by a uniform dilatation, because it is energetically more convenient to create a cavity (see Fig. \ref{rubber}(c)). Although this creates a point defect in the reference configuration at which $W$ is infinite, the overall energy is decreased because $\int_\om h(\det\nabla\y)\,d\x$ is much smaller. This is rigorously analyzed in the class of radial deformations in \cite{j19}. In fact cavitation is a standard failure mechanism for rubber (see, for example, \cite{gent58,lazzeri95}).

However, maps $\y\in H^1(\om;\R^3)$ cannot have planar discontinuities such as occur in cracks. In order to model such fracture we must enlarge the function space again, this time to be the space $SBV(\om;\R^3)$ consisting of special mappings of bounded variation \cite{ambrosioetal00}. This space is somewhat technical to describe, but the key points are that (i) $\y\in SBV(\om;\R^3)$ can have jump discontinuities at points of a well-defined {\it jump set} $S_\y$, at which there is a well-defined unit normal $\nnu$ with respect to which there are well-defined limiting values $\y_+, \y_-$ from either side of $S_\y$, (ii) the gradient $\nabla\y$ can still be defined away from $S_\y$. There is, however, an energetic penalty $f(\y_+,\y_-,\mathbf\nu)$ per unit area associated with a jump in $\y$, so that the energy functional has to be modified to
\be 
\label{energyfracture}
\hat I(\y)=\int_\om W(\nabla\y)\,d\x+\int_{S_\y}f(\y_+,\y_-,\nnu)\,d\mH^2,
\ee
for some suitable $f$, where $d\mH^2$ denotes the element of area on $S_\y$. This functional coincides with $I(\y)$ for $\y\in H^1(\om;\R^3)$. Then we expect to obtain a different minimizer with cracks (see Fig. \ref{rubber}(d)). Such `free discontinuity' models for fracture were first proposed by Francfort \& Marigo \cite{francfortmarigo98}.  For a rigorous treatment of fracture in which cavitation is also allowed, together with numerical computations,  see   \cite{HenaoMoraCorral2010,HenaoMoraCorral2011,HenaoMoraCorralXu2015,HenaoMoraCorralXu2016}. In summary, we have an energy functional $\hat I$ having different minimizers in three different function spaces. 

\subsection{The Lavrentiev phenomenon for liquid crystals}
\label{lav}

But does the Lavrentiev phenomenon occur for liquid crystals? As a first example we can consider the hedgehog $\hat\n$ defined in \eqref{hedgehog}. Since for topological reasons there is no continuous $\n$ satisfying $\n=\hat\n$ on the boundary $\partial B$ of the unit ball $B$, we have formally that 
\be
\label{nocontinuous}
I(\n)=\int_BW(\n,\nabla\n)\,d\x=\infty \hspace{.1in}\mbox{for all continuous } \n \mbox{ with } \n|_{\partial B}=\hat\n|_{\partial B},
\ee 
whereas $I(\hat\n)<\infty$. 

Lest \eqref{nocontinuous} be regarded as somehow cheating, Hardt \& Lin \cite{hardtlin} constructed an example for the one-constant approximation in which there is 
a  smooth (degree zero) map 
$\bar\n:B\to S^2$ such that for some $\alpha>0$
$$\min_{\n\in H^1(B;S^2), \n|_{\partial B}=\bar\n}I(\n)+\alpha< I(\m)$$
for all continuous $\m\in H^1(B;S^2)$. In this case there are smooth maps for which the energy is finite, but they all have energy at least $\alpha$ greater than the minimum among $H^1(B;S^2)$ maps.

A more interesting possibility is if we allow the director $\n$ to be discontinuous across surfaces (planar defects), in the same spirit as the fracture models of Francfort \& Marigo, so that we have an augmented functional defined for 
$\n\in SBV(\om;S^2)$ by
\be 
\label{newfunctional}
\hat I(\n)=\int_\Omega W(\n,\nabla\n)\,d\x+\int_{S_\n}f(\n_+,\n_-,{\nnu})\,d\mH^2,
\ee
for some suitable continuous interfacial energy function $f$. As before, $S_\n$ denotes the jump set of $\n$, $\nnu$ its unit normal, and $\n_+, \n_-$ the limiting values of $\n$ from either side of $S_\n$. Such models have been investigated in \cite{u14,j67}, where in particular it is shown that invariance requirements imply that $f$ depends only on the quantities $(\n_+\cdot\n_-)^2, (\n_+\cdot\nnu)^2, (\n_-\cdot\nnu)^2$ and $(\n_+\cdot\n_-)(\n_+\cdot\nnu)(\n_-\cdot\nnu)$. See also Bedford \cite{bedfordcholesterics}.

To see why the Lavrentiev phenomenon arises for $\hat I$,  consider the order reconstruction problem in which a nematic liquid crystal occupies the region  $\om=(0,l_1)\times(0,l_2)\times (0,d)$ between two parallel plates given by $x_3=0$ and $x_3=d$, where $d>0$, with antagonistic boundary conditions on the plates
\be 
\label{antagonistic}
\n(x_1,x_2,0)= \e_1,\;\;\n(x_1,x_2,d)=\e_3, \mbox{ for }(x_1,x_2)\in (0,l_1)\times(0,l_2),
\ee
and periodic boundary conditions on the other faces 
\be 
\label{periodic}
\n(0,x_2,x_3)=\n(l_1,x_2,x_3),\;\; \n(x_1,0,x_3)=\n(x_1,l_2,x_3).
\ee
Such problems have been considered by many authors  using   the Landau - de Gennes model (see Section \ref{ldg}) \cite{barberietal04,barberobarberi,bisietal03,carboneetal,lamy14,palffyetal},  the variable scalar order parameter model of Ericksen \cite{ambrosiovirga91}, and molecular dynamics \cite{zannonithinfilm}. If $\n\in H^1(\om; S^2)$   satisfies the boundary conditions \eqref{antagonistic}, \eqref{periodic}, then by \eqref{lowerbound} and Jensen's inequality $\int_\om|\vu|^2d\x\geq |\om|^{-1}\left|\int_\om \vu\,d\x\right|^2$, where $|\om|=l_1l_2d$ denotes the volume of $\om$, we have that
\be 
\hat I(\n)&=&\int_\om W(\n,\nabla\n)\,d\x \nonumber\\
&\geq&c_0\int_\om |\nabla\n|^2d\x\nonumber\\
&\geq& c_0|\om|^{-1}\left|\int_\om \nabla\n\,d\x\right|^2\nonumber\\
&=&c_0|\om|^{-1}(l_1l_2)^2|\e_3-\e_1|^2\nonumber\\
&=&2c_0 \frac{l_1l_2}{d}.\label{est}
\ee
However, in the space $SBV(\om;S^2)$ we can consider the competitor
\be 
\label{jumpsolution}
\N=\left\{\begin{array}{ll} \e_1,& 0<x_3<\frac{d}{2}\\
\e_3,&\frac{d}{2}<x_3\leq d \end{array}\right.,
\ee 
for which
\be 
\label{est1}
\hat I(\N)=l_1l_2f(\e_1,\e_3,\e_3),
\ee 
which is less than $2c_0 \frac{l_1l_2}{d}$ provided 
\be 
\label{est2}
d<\frac{2c_0}{f(\e_1,\e_3,\e_3)}.
\ee
Thus for sufficiently small plate separation the minimum value for $\hat I$ in the space $SBV(\om;S^2)$ is less than that in $H^1(\om;S^2)$. For more details see \cite{u14,j67}.

The use of $SBV(\om;S^2)$ also has the purely mathematical advantage of treating index $\frac{1}{2}$ defects using a vector field rather than a line-field, by allowing $\n$ to jump to $-\n$ across surfaces with zero energy cost, such as  the surface OP indicated in Fig. \ref{indexhalf}(c). This is explored in Bedford \cite{bedfordcholesterics}. In \cite{j67} further possible applications to stripe domains in nematic elastomers and smectic thin films are suggested.

\section{The Landau - de Gennes model}
\label{ldg}
The Landau - de Gennes model uses a five-dimensional tensor order parameter based on the probability 
distribution $\rho(\x,\p)$ of molecular orientations $\p\in S^2$ at a point $\x$. Here $\p$ is parallel to the long axis of a molecule, and we regard $\p$ and $-\p$ as being equivalent. Two generally perceived 
advantages of the Landau - de Gennes model over that of Oseen-Frank  are that  (i) it gives structure to
defects, so that in particular they have finite energy, (ii) it resolves 
the problem of orientability of the director.

\subsection{The probability distribution of molecular orientations}
\label{probmol}
In order to give meaning to the probability distribution $\rho(\x,\p)$, consider   the ball $B(\x,\delta)=\{\z\in \R^3: |\z-\x|<\delta\}$ with centre $\x$ and small radius $\delta>0$. We 
want $\delta$ to be smaller than macroscopic length scales, but large enough to contain many molecules. To give an idea of the numbers, if $\delta=1\mu$m then $B(\x,\delta)$ would contain about a billion molecules. Then $\rho(\x,\p)$ can be thought of as (a smoothed out version of) the probability of a molecule chosen at random from those in $B(\x,\delta)$ having orientation $\p$. See Section \ref{molmodels} for further discussion. 

Since $\rho$ is a probability distribution and $\pm\p$ are equivalent, $\rho$ satisfies
\be 
\label{rhoprops}
\rho(\x,\p)\geq 0, \;\;
\rho(\x,\p)=\rho(\x,-\p), \;\;
\int_{S^2}\rho(\x,\p)\,d\p=1,
\ee
where $d\p$ denotes the area element on $S^2$. By \eqref{rhoprops} the first moment 
\be 
\label{firstmoment}\int_{S^2}\p\,\rho(\x,\p)\,d\p=-\int_{S^2}\p\,\rho(\x,-\p)\,d\p=\bzero.
\ee 
The second moment
\be 
\label{second}
{\M}(\x)=\int_{S^2}\p\otimes\p\,\rho(\x,\p)\,d\p 
\ee
is a symmetric  second-order tensor which is positive definite, that is $\M\e\cdot\e>0$ for all $\e\in S^2$. Indeed
\be 
\label{posmatrix}
{\M}(\x)\e\cdot\e=\int_{S^2}(\p\cdot\e)^2\,\rho(\x,\p)\,d\p\geq 0,
\ee
while $\M(\x)\e\cdot\e=0$ implies that $\rho(\x,\p)=0$ for $\p\cdot\e\neq 0$, contradicting $\int_{S^2}\rho(\x,\p)\,d\p=1$. Also $\tr \M(\x)=1$.
The case $\rho(\x,\p)=\frac{1}{4\pi}$ corresponds to an {\it isotropic} molecular distribution at $\x$, and it is easily checked that then  ${\M}(\x)=\frac{1}{3}{\bf 1}$.

The de Gennes $\Q$ tensor is defined as
\begin{eqnarray}\label{Qtensor}
\Q(\x)&=&{\mathbf M}(\x)-\frac{1}{3}{\bf 1}\\
&=&\int_{S^2}(\p\otimes \p-\frac{1}{3}{\bf 1})\rho(\x,\p)\,d\p\nonumber.
\end{eqnarray} 
Thus $\Q(\x)$ measures the deviation of $\M(\x)$ from its isotropic value, and
 from the properties of $\M(\x)$ satisfies 
\be 
\label{Qprops}
\Q(\x)=\Q^T(\x),\;\tr \Q(\x)=0,\, \lambda_{\rm min}(\Q(\x))>-\frac{1}{3},
\ee
where $\lambda_{\rm min}(\Q(\x))$ denotes the minimum eigenvalue of $\Q(\x)$. Thus $\Q(\x)$ is a five-dimensional order parameter, whereas the director $\n(\x)$ is two-dimensional.

\subsection{The Landau - de Gennes energy functional}
Following de Gennes we suppose that the free energy for a nematic at constant temperature is given by
\be 
\label{Qfree}
I(\Q)=\int_\Omega \psi(\Q,\nabla\Q)\,d\x.
\ee
Writing $\vH=\nabla\Q$, where $\Q=(Q_{ij}), \vH=(H_{ijk})=(Q_{ij,k})$, as a consequence of frame-indifference and material symmetry,   $\psi=\psi(\Q,\vH)$ should satisfy for any $\vR=(R_{ij})\in O(3)=\{\vR:\vR^T\vR=\1\}$ the isotropy condition
\be 
\label{isotropy}
\psi(\Q^*,\vH^*)=\psi(\Q,\vH),
\ee
where $Q^*_{ij}=R_{ir}R_{js}Q_{rs}, H^*_{ijk}=R_{ir}R_{js}R_{kt}H_{rst}$.

It is usual to decompose the free-energy density $\psi$ as
\begin{eqnarray}
\psi(\Q,\nabla \Q)&=&\psi(\Q,{\mathbf 0})+(\psi(\Q,\nabla \Q)-\psi(\Q,{\mathbf 0}))\nonumber\\
&=&\psi_B(\Q)+\psi_E(\Q,\nabla \Q)\label{decomp}\\
&=& \mbox{ bulk }+\mbox{ elastic}.\nonumber
\end{eqnarray}
It is often assumed that $\psi_B$ has the form
\be 
\label{psiB}
\psi_B(\Q)=a{\rm tr}\,\Q^2-\frac{2b}{3}{\rm tr}\,\Q^3+c{\rm tr}\,\Q^4,
\ee
where $b>0,c>0$   and $a$ depends linearly on temperature. If $a<\frac{b^2}{27c}$ then $\psi_B$ is minimized
by $\Q$ having the {\it uniaxial} form 
\be 
\label{uni}\Q=s(\n\otimes\n-\frac{1}{3}{\bf 1}),\; \n\in S^2,
\ee
where 
\be 
\label{s}
s=\frac{b+\sqrt{b^2-24ac}}{4c}>0. 
\ee

Usually it is assumed that $\psi_E(\Q,\nabla \Q)$ is quadratic in $\nabla \Q$.
Examples of isotropic functions quadratic  in $\nabla \Q$ are the invariants $I_i=I_i(\Q,\nabla\Q)$:
\begin{eqnarray}
     &&I_1 =   Q_{ij,k}Q_{ij,k}, \;\;
    I_2 =   Q_{ij,j}  Q_{ik,k}, \nonumber\\
   &&I_3 =  Q_{ik,j}Q_{ij,k}, \;\;
    I_4 = Q_{lk} Q_{ij,l} Q_{ij,k}.\label{isotr}
\end{eqnarray}
The first three linearly independent invariants $I_1,I_2,I_3$ span the possible isotropic quadratic functions of $\nabla\Q$.  The invariant $I_4$ is one of 6 possible linearly independent cubic terms that are quadratic in $\nabla \Q$ (see  \cite{berremanmeiboom1984,longaetal1987,poniewierskisluckin1985,schieletrimper1983}). Note that
\be 
\label{nullL}
I_2-I_3=(Q_{ij}Q_{ik,k})_{,j}-(Q_{ij}Q_{ik,j})_{,k}
\ee
 is a null Lagrangian.

We assume that 
\be 
\label{elasticform}
\psi_E(\Q,\nabla\Q)=\sum_{i=1}^4L_iI_i,
\ee
where the $L_i$ are material constants\footnote{Here we use the definitions of the $L_i, i=1,2,3$ most common in the literature (for example \cite{degennes1971,longaetal1987,morietal}), rather than those used in the 
 papers \cite{j67,j59,j57,j61} in which the   $L_i$ were permuted with respect to these definitions, with a corresponding permutation of the $I_i$.}.
Note that if 
$L_2=L_3=L_4=0$ then $\psi_E=L_1|\nabla\Q|^2$, the {\it one-constant approximation} of the Landau - de Gennes theory.

\subsection{From Landau - de Gennes to Oseen-Frank}
Since $\psi_B$ is minimized for uniaxial $\Q$ given by (see \eqref{uni})
\be 
\label{uni1}\Q=s(\n\otimes\n-\frac{1}{3}{\bf 1}),\; \n\in S^2,
\ee
in the limit of
small elastic constants $L_i$ (see below for further discussion) we expect minimizers of 
$I(\Q)$ to be nearly uniaxial. This motivates the 
{\it constrained theory}\footnote{For a corresponding constrained theory for biaxial nematics see \cite{muccinicolodi2012,muccinicolodi2016}.} in which we minimize $I(\Q)$ subject to the constraint \eqref{uni1} for fixed $s>0$.
 Putting $\eqref{uni}$ into $\psi_E$ we obtain the Oseen-Frank
energy
 \begin{eqnarray*} \nonumber  &&\hspace{-.3in}W(\n,\nabla\n) = K_{1}(\textrm{div}\,
\n)^{2} + K_{2}(\n\cdot \textrm{curl}\,\n)^{2}  +
K_{3}|\n \wedge \textrm{curl}\,\n|^{2}\\&&\hspace{2in}  +
(K_{2}+K_{4})(\textrm{tr}(\nabla \n)^{2} - (\textrm{div}\,
\n)^{2}),
\end{eqnarray*}
with
\begin{eqnarray}\label{KtoL}
\left(
\begin{array}{c}K_1\\ K_2\\ K_3\\ K_4\end{array}
\right)=s^2\left(
\begin{array}{cccc}2&1&1&-\frac{2}{3}s\\
2&0&0&-\frac{2}{3}s\\ 2&1&1&\frac{4}{3}s\\
0&0&1&0\end{array}
\right)\left(
\begin{array}{c}L_1\\ L_2\\ L_3\\ L_4\end{array}
\right).
\end{eqnarray}
The $4\times 4$ matrix in \eqref{KtoL} is invertible, so that the $L_i$ can be also be obtained from the $K_i$ through the formula
\begin{eqnarray}\label{KtoLinv}
\left(
\begin{array}{c}L_1\\ L_2\\ L_3\\ L_4\end{array}
\right)=s^{-2}\left(
\begin{array}{cccc}-\frac{1}{6}&\frac{1}{2}&\frac{1}{6}&0\\1&-1&0&-1\\0&0&0&1\\
-\frac{1}{2s}&0&\frac{1}{2s}&0\end{array}
\right)\left(
\begin{array}{c}K_1\\ K_2\\ K_3\\ K_4\end{array}
\right).
\end{eqnarray}
 This is one reason why we choose the simplified form \eqref{elasticform} for the elastic part of the energy, rather than including all the possible cubic terms.  Including $I_4$ as the single cubic term   is a common choice (see, for example, \cite{mottramnewton,ravnikzumer2009}), and is convenient for establishing existence when $L_4\neq 0$ (see the discussion at the end of Section \ref{singbulk}).

As observed by Gartland \cite{gartland2015}, care needs to be taken as regards the physical interpretation of the $L_i$ being small, because their values depend on the units of length used. A change of units of length can be represented by the change of variables $\x=\lambda \y, \lambda>0$, $\tilde \Q(\y)=\Q(\lambda\y)$, leading to $I(\Q)=\tilde I(\tilde\Q)$, where
\be 
\label{newIa}
\tilde I(\tilde\Q)=\lambda^3\int_{\om/\lambda}\left(\psi_B(\tilde\Q)+\frac{1}{\lambda^2}\sum_{i=1}^4L_iI_i(\tilde\Q,\nabla\tilde\Q)\right)\,d\y.
\ee
If the diameter of the physical domain $\om$ in the $\x-$coordinates is $R$, and we set $\lambda=R$ then we can write $\om=\lambda\tilde \om$ with $\diam\tilde\om=1$. Writing $\psi_B=e_0\tilde\psi_B$, where $e_0$ has dimensions of energy per unit volume,    \eqref{newIa} takes the form
\be 
\label{nnewI}
\tilde I(\tilde\Q)=R^3e_0\int_{\tilde\om}\left(\tilde\psi_B(\tilde\Q)+\sum_{i=1}^4\frac{L_i}{e_0R^2}I_i(\tilde\Q,\nabla\tilde\Q)\right)\,d\y,
\ee
in which the scaled  coefficients $L_i'=\frac{L_i}{e_0R^2}$ are dimensionless. The constrained theory should be applicable when the $L_i'$ are small, which in particular is the case in the large body limit $R\to\infty$ for {\it fixed} $L_i$.

There are a number of rigorous results concerning the passage from the Landau - de Gennes model to that of Oseen-Frank.  Most are for the one-constant approximation
\be 
\label{onec}
I_L(\Q)=\int_\Omega(\psi_B(\Q)+L|\nabla \Q|^2)\,d\x
\ee 
with prescribed uniaxial boundary data,  in the limit as $L\to 0+$, with $\psi_B$ given by \eqref{psiB} with
 $a<0$. 

 Majumdar \& Zarnescu \cite{majumdarzarnescu} showed that 
for any sequence  $L^{(k)}\to 0+$  there is a subsequence $L^{(k')}$
such that minimizers $\Q^{(k')}$ for $I_{L^{(k')}}$ converge
in $H^1(\om;\R^3)$ to a minimizer $\Q^*=s(\n^*\otimes \n^*-\frac{1}{3}{\bf 1})$ of $\int_\Omega |\nabla\Q|^2d\x$
subject to the constraint $\Q=s(\n\otimes\n-\frac{1}{3}{\bf 1})$, the convergence meaning that  $\int_\om|\Q^{(k')}-\Q^*|^2d\x\to 0$ and $\int_\om|\nabla\Q^{(k')}-\nabla\Q^*|^2d\x\to 0$.
If $\Omega$ is simply-connected then 
 any uniaxial $\Q\in H^1(\om;\R^3)$ is orientable \cite{j61}. Hence $\n^*$ is a harmonic map
(a minimizer for the one-constant Oseen-Frank energy), and so by Theorem \ref{finitenumber} has a finite number of point defects.
Away from these defects the convergence $\Q^{(k')}\to\Q^*$
is much better.   The rate of convergence   was improved by Nguyen \& Zarnescu \cite{nguyenzarnescu}, who also obtained a first-order correction term.

 Canevari \cite{Canevari2016} studied the convergence of minimizers
$\Q_L$ as $L\to 0+$ under the logarithmic scaling
\be 
\label{logscaling}
I_L(\Q_L)\leq CL|\ln L|,
\ee 
which allows the appearance of line defects in the limit. He shows
that these defects consist of {\it straight line segments}. 
 (For colloids, where curved disclinations are seen, there are other
small geometric parameters. See the study of Saturn rings by
Alama, Bronsard \& Lamy  \cite{AlamaBronsardLamy2016}.) In earlier work, Bauman, Park \& Phillips \cite{baumanparkphillips2012} studied how line defects appeared  in the zero elastic constant limit for thin films, for general elastic constants $K_1,K_2, K_3$ with $K_4=0$. 

 Given the appearance of uniaxial $\Q$ in the limit $L\to 0$ one
can ask whether in fact there are equilibrium solutions for
$I_L(\Q)$ with $L>0$ which are {\it everywhere uniaxial} with director $\n$ and $s=s(\x)$ (the 
starting assumption of the  Ericksen theory of liquid crystals \cite{ericksen1991liquid}). Lamy \cite{Lamy2015} shows that in one and 
two dimensional situations $\n$ has to be constant. These results suggest that, although for small $L$ configurations predicted by Landau - de Gennes
are very close to being uniaxial (at least away from defects), they 
are rarely exactly so.

\subsection{Description of defects in the Landau - de Gennes model}
 In contrast to the Oseen-Frank theory, we expect minimizers
of $I(\Q)$ subject to suitable boundary conditions, 
such as $\Q|_{\partial\Omega}=\bar\Q$ where $\bar\Q$ is given, to be {\it smooth}.
When $L_4=0$ this was proved by 
Davis \& Gartland \cite{gartlanddavis} under the conditions
\be 
\label{Lcondns}L_1>0, -L_1<L_3<2L_1,  L_1+\frac{5}{3}L_2+\frac{1}{6}L_3>0,
\ee
which imply   \cite{longaetal1987} that $\psi_E(\Q,\nabla\Q)\geq c|\nabla \Q|^2$ for some $c>0$,
so that $\psi_E(\Q,\cdot)$ is convex, and the corresponding Euler-Lagrange equations 
form a semilinear elliptic system.

Thus defects in the Landau - de Gennes theory {\it are not described 
by singularities in $\Q$.} This has been explored for the hedgehog defect 
by many authors (see, for example, \cite{GartlandMkaddem1999,HenaoMajumdar2012,HenaoMajumdarPisante2016,ignatetal2015,KraljVirga2001,lamy2013,Majumdar2012}), and for defects of other index in \cite{ignatetal2015a,ignatetal2016}.

However $L_4=0$ implies $K_1=K_3$, so we need $L_4\neq 0$.  
But then 
\begin{thm}[Ball \& Majumdar \cite{u9,j59}] 
\label{unbounded}If $L_4\neq 0$, and for any continuous 
$\psi_B$ and any boundary conditions, $$I(\Q)=\int_\Omega
(\psi_B(\Q)+\psi_E(\Q,\nabla\Q))\,d\x$$ is {\it unbounded below.}
\end{thm}

\section{Onsager and molecular models}
\label{onsagermol}
\subsection{Onsager models}
In the Onsager model, the free energy for a homogeneous nematic liquid crystal at temperature $\theta>0$ is given in terms of the probability distribution $\rho=\rho(\p)$ of molecular orientations by
\be 
\label{onsager}I_\theta(\rho)=k_B\theta\int_{S^2}\rho(\p) \ln\rho(\p)\,d\p+\half\int_{S^2}\int_{S^2}k(\p\cdot\q)\rho(\p)\rho(\q)\,d\p\, d\q.
\ee
As in Section \ref{ldg}, we assume that $\rho$ satisfies the condition of head-to-tail symmetry 
\be 
\label{htt}
\rho(\p)=\rho(-\p) \mbox{  for all } \p\in S^2.
\ee
 Two well-known examples of suitable kernels $k=k(\p\cdot\q)$ are\vspace{.1in}\\
(i) $k(\p\cdot \q)=\kappa(\frac{1}{3}-(\p\cdot \q)^2)$\hspace{.5in} (Maier-Saupe),\vspace{.05in}\\ 
(ii) $k(\p\cdot \q)=\kappa \sqrt{1-(\p\cdot \q)^2}$ \hspace{.5in}(Onsager),\vspace{.05in}\\ 
where $\kappa>0$ is a  constant, which may depend on temperature and concentration. For both of these cases the behaviour of solutions depends only on the dimensionless parameter
\be 
\label{alpha}
\alpha=\frac{\kappa}{k_B\theta}.
\ee

The Euler-Lagrange equation corresponding to the free-energy functional \eqref{onsager} subject to the constraint $\int_{S^2}\rho(\p)\,d\p=1$ is 
\be 
\label{ELonsager}
k_B\theta\ln\rho(\p)+\int_{S^2}k(\p\cdot\q)\rho(\q)\,d\q= c,
\ee
where $c$ is a constant Lagrange multiplier, for which one solution valid for all $\theta>0$ is  the isotropic state
\be 
\label{isotropicstate}
\bar\rho(\p)=\frac{1}{4\pi}.
\ee 
 Note that if $\rho=\rho(\p)$ is a solution, so is $\rho_\vR(\p)=\rho(\vR\p)$ for any $\vR\in SO(3)$. In order to study all possible solutions one can begin by looking for solutions bifurcating from $\bar\rho$. The possible bifurcation points correspond to nonzero solution pairs $(\varphi,\lambda)$ with $\int_{S^2}\varphi(\p)\,d\p=0$  to the equation obtained by linearizing \eqref{ELonsager} about $\bar\rho$, namely
\be 
\label{bifeqn}
\int_{S^2}k(\p\cdot\q)\varphi(\q)\,d\q=\lambda\varphi(\p),
\ee  
where $\lambda=-4\pi k_B\theta$.
The eigenvalues $\lambda<0$ of \eqref{bifeqn} have been calculated explicitly for a wide range of kernels $k$ by Vollmer \cite{Vollmer2016}. The case of the Maier-Saupe kernel is unusual in that there is only one such eigenvalue $\lambda^*=-\frac{8\pi}{15}\kappa$, corresponding to $\alpha= \frac{15}{2}$ in \eqref{alpha}, where there is a transcritical bifurcation,  and for this kernel all critical points can be explicitly determined (Fatkullin \& Slastikov \cite{FatkullinSlastikov2005}, Liu, Zhang \& Zhang \cite{LiuZhangZhang2005}).

For the Onsager kernel, there are infinitely many distinct bifurcation points $\alpha_j$.  Vollmer \cite{Vollmer2016} (see also Wachsmuth \cite{wachsmuth}) shows that at the least bifurcation point $\alpha=\frac{32}{\pi}$ there is a transcritical bifurcation to   an axially symmetric state, together with rotations of it, and proves other properties of the solution set. However a complete description of the set of all solutions remains an open problem. For the 2D case see Lucia \& Vukadinovic \cite{luciavukadinovic2010}. 

\subsection{The singular bulk potential.}
\label{singbulk}
The $\Q$ tensor corresponding to $\rho$ is
 \be 
\label{Qrho}
\Q(\rho)=\int_{S^2}(\p\otimes \p-\frac{1}{3}{\bf 1})\rho(\p)\,d\p.
\ee
Following Katriel, Kventsel, Luckhurst \& Sluckin \cite{katrieletal} and Ball \& Majumdar \cite{u9,j59},
the bulk energy $\psi_B(\Q,\theta)$ can  be identified with
the minimum of $I_\theta(\rho)$ among probability distributions 
$\rho$ on $S^2$ satisfying \eqref{htt} such that $\Q(\rho)=\Q$.

For the Maier-Saupe kernel the interaction 
term takes the simple form
\be 
\label{MSform}
\frac{\kappa}{2}\int_{S^2}\int_{S^2}\left(\frac{1}{3}-(\p\cdot \q)^2\right)\rho(\p)\rho(\q)\,d\p\,d\q=-\frac{\kappa}{2}|\Q(\rho)|^2,
\ee
so that 
\be 
\label{energyMS}
I_\theta(\rho)=k_B\theta\int_{S^2}\rho(\p)\ln\rho(\p)\,d\p-\frac{\kappa}{2}|\Q(\rho)|^2.
\ee
Hence 
\be 
\label{psiB1}
\psi_B(\Q,\theta)=k_B\theta f(\Q)-\frac{\kappa}{2}|\Q|^2,
\ee
 where
\be 
\label{deff}
f(\Q)=\min_{\rho, \Q(\rho)=\Q}\int_{S^2}\rho(\p)\ln\rho(\p)\,d\p.
\ee

\begin{thm}[\cite{u9,j59}]\label{logest1} $f$ is a smooth strictly convex
function of $\Q$ defined for $\lambda_{\rm min}(\Q)>-\frac{1}{3}$, and 
\be 
\label{logest}
C_1-\frac{1}{2}\ln (\lambda_{\rm min}(\Q)+\frac{1}{3})\leq f(\Q)\leq C_2- \ln (\lambda_{\rm min}(\Q)+\frac{1}{3}),
\ee
for constants $C_1, C_2$.
\end{thm}

 \noindent In particular the {\it singular potential} $\psi_B^s(\Q,\theta)=\psi_B(\Q,\theta)$  
satisfies 
\be 
\label{blowup}\psi_B^s(\Q,\theta)\to \infty \mbox{ as } \lambda_{\rm min}(\Q)\to -\frac{1}{3}+.
\ee
 Using this bulk potential, under suitable conditions on the $L_i$ it follows that the energy $I(\Q)$ in \eqref{Qfree} 
is bounded below and attains a minimum even for $L_4\neq 0$, in contrast to Theorem \ref{unbounded}. To see this, note that $\lambda_{\rm min}(\Q)\geq -\frac{1}{3}$ implies that
\be 
\label{I4I3}
-\frac{1}{3}I_1\leq I_4\leq\frac{2}{3}I_1.
\ee
Define
\be 
\label{newL3}
L_1'=\left\{\begin{array}{ll}L_1-\frac{1}{3}L_4&\mbox{if }L_4\geq 0\\L_1+\frac{2}{3}L_4&\mbox{if }L_4\leq 0\end{array}\right. .
\ee
Thus by \eqref{Lcondns} if
\be 
\label{newLcondns}
L_1'>0, -L_1'<L_3<2L_1', L_1'+\frac{5}{3}L_2+\frac{1}{6}L_3>0,
\ee
we have that $\psi_E(\Q,\nabla\Q)\geq c_1|\nabla\Q|^2$ for some $c_1>0$, implying in particular that the $K_i$ given in  \eqref{KtoL} satisfy the Ericksen inequalities \eqref{ericksen}, as can be checked directly.

Furthermore the singular potential predicts an isotropic to nematic phase transition in the same way as the quartic potential \eqref{psiB}, which can be regarded as an expansion of the singular potential around the isotropic state $\Q=\bzero$. 

 For a general framework for singular potentials together with applications see the recent work of Taylor \cite{Taylor2016}.

\subsection{Molecular models}
\label{molmodels}
 It is an important open problem to establish some kind
of rigorous link between molecular models of liquid crystals, however primitive,
and continuum models, so that solutions to the equations of motion for the molecules are shown to converge in an appropriate average sense to those for some continuum model. This is a deep problem even for the much simpler situation of a moderately rarified gas, for which there is a large literature (see, for example,   \cite{GallagherSaintRaymondTexier2013,Lanford1975}) on the passage from Newtonian mechanics of hard spheres to a   description of the dynamics of the gas in terms of a probability distribution for the velocities satisfying the Boltzmann equation, and a further large literature (see, for example, \cite{GorbanKarlin2014,SaintRaymond2014}) on how to pass from the Boltzmann equation to the equations of continuum fluid dynamics. There do not seem to be any corresponding rigorous results known for liquid crystals, though there are  penetrating formal studies (for example, 
\cite{GelbartBenShaul1982,NehringSaupe1971}).

One issue is how the macroscopic variables,
such as $\rho(\p)$ and its moments, would emerge in such a theoretical framework. We already gave in Section \ref{probmol} a suggestive description of how this might happen for nematics via a coarse-graining procedure. Less clear is the situation for smectics, in  
  some theories of which (for example \cite{chenlubensky,degennes72,hanetal,klemanparodi,mcmillan,selingersluckin}) the molecular mass density
$c(\x)$ plays a role,   its variation describing smectic layers.
How can we have a macroscopic variable which varies on a 
microscopic length scale?  One might try to resolve this 
by averaging at a fixed time over a ball centred at $\x$ and small radius $\delta$ sufficient to contain
lots of molecules. However, it is not obvious this works because such an
average may not detect oscillations. For example,
\be 
\label{sin}
\frac{1}{\mbox{vol}\,B(\x,\delta)}\int_{B(\x,\delta)}\sin(kz_1)\,dz
=3\sin(kx_1)
\left(\frac{\sin k\delta-k\delta\cos k\delta}{(k\delta)^3}\right),
\ee 
is zero for all $\x$ if $k\delta$ is a root of the equation  $\sin \lambda=\lambda\cos\lambda.$
 This is related to  the {\it Pompeiu problem} \cite{agranovich2008,ramm2005}, which asks for which bounded domains $\Omega\subset\R^n$ is it true that the only locally integrable function $f:\R^n\to\R$ for which
$$\int_{\Rz\om+\az}f(\x)\,d\x=0 \mbox{ for all } \Rz\in SO(n), \az\in\R^n,$$
is $f=0$,  for which \eqref{sin} is a counterexample for $\Omega$ a ball. A possible remedy would be to average also over a short time interval, so that the radius $\delta$ could be made smaller than the layer thickness. To make this rigorous would seem to require a quite detailed understanding of the molecular motion in smectics. 
\section{Unequal elastic constants}
\label{unequal}
We have seen that much rigorous mathematical work on liquid crystals
makes the one-constant approximation (both for Oseen-Frank and Landau - de Gennes). In this section we highlight some rigorous results for the case of unequal elastic constants.
\subsection {Oseen-Frank}
 One reason why the case of general $K_i$ may be more difficult is that it
is only for the one-constant approximation that the Lagrange multiplier
$\lambda(x)$ in the Euler-Lagrange equation \eqref{lagmult}, that is
$${\rm div}\,\frac{\partial W}{\partial \nabla \n}-\frac{\partial W}{\partial \n}=\lambda(\x)\n,$$
depends only on $\nabla\n$.

In the one-constant approximation (Theorem \ref{finitenumber}) energy minimizers are smooth except for
a finite number of point defects. This is not known for general
$K_i$. However, we have

\begin{thm} [Hardt, Lin \& Kinderlehrer \cite{hardtkinderlehrerlin}] Let $K_1>0, K_2>0, K_3>0$, and let   $\bar\n\in H^1(\Omega;S^2)$.  Then any minimizer $\n\in H^1(\Omega;S^2)$ of
$$I(\n)=\int_\om W(\n,\nabla\n)\,d\x$$
subject to $\n|_{\partial\om}=\bar\n$, where $W$ is given by \eqref{OFE},
is analytic outside a closed subset $\mathcal S$ of $\om$ whose Hausdorff dimension is less than one.
\end{thm}
\noindent Hardt, Lin \& Kinderlehrer \cite{hardtkinderlehrerlin} also prove a corresponding partial regularity result up to the boundary.

An extension of Rivi\`ere's result for the one-constant approximation to general $K_i$ is
due to Hong \cite{hong2007}, who proves that if 
$$|K_1-K_2|\leq 4\left(\frac{1-\ln 2}{\ln 2}\right)\min (K_1,K_2),$$
 and if $\partial\Omega$ is smooth with $\bar\n$ smooth on $\partial\om$ and nonconstant, then there are infinitely 
many solutions to $(WEL)$ satisfying $\n|_{\partial\Omega}=\bar\n$  (see  Section \ref{enmin} for the special case of the hedgehog on a ball).

\subsubsection{Energy minimizing properties of universal solutions}
\label{enmin}
It is interesting to ask under which conditions on the $K_i$, and for which boundary conditions, the universal solutions described in Section \ref{universal} are energy minimizers. For the hedgehog $\hat\n(\x)=\frac{\x}{|\x|}$, it was observed by H\'elein \cite{helein} that the proof of Lin \cite{lin1987}  works in the case $K_2\geq K_1>0, K_3>0$, so that under these conditions on the $K_i$ the hedgehog minimizes $I(\n)$ subject to its own boundary conditions. A detailed proof of this for $\Omega$ a ball is given in \cite{ou92}; for a general bounded domain $\om$ one can extend any $\n\in H^1(\om;S^2)$ with $\n|_{\partial\om}=\hat\n$ to a larger ball $B$ by setting $\n=\hat\n$ outside $\om$, and then apply the result for $B$, but a direct proof is also possible. The hedgehog is a pure splay configuration, since $\curl\hat\n=0$. The condition that $K_2\geq K_1$ says that it is energetically easier to splay than to twist. 

Necessary and sufficient conditions for $\hat\n$ to be a minimizer subject to its own boundary conditions do not seem to be known. H\'elein \cite{helein} showed that if $8(K_2-K_1)+K_3<0$ then the second (outer) variation of $I(\n)$ at $\hat\n$ can be negative, so that $\hat\n$ is not a minimizer. On the other hand Cohen \& Taylor \cite{cohentaylor} showed that if the opposite inequality $8(K_2-K_1)+K_3\geq 0$ holds then the second variation of $I(\n)$ at $\hat\n$ is strictly positive, so that with respect to certain variations $\hat\n$ is a local minimizer. A simplified proof of these second variation calculations was given by Kinderlehrer \& Ou \cite{KinderlehrerOu1992}. Alouges \& Ghidaglia \cite{AlougesGhidaglia1997} presented numerical computations indicating that the condition $8(K_2-K_1)+K_3\geq 0$ is not sufficient for $\hat\n$ to be a global minimizer; however, $K_1=K_2$ for two of the three cases when a configuration with lower energy was apparently found, when by the result mentioned  above $\hat\n$ is in fact a minimizer.

Hong \cite{hong04} shows that for a ball $B$  there are always at least two solutions of $(WEL)$ satisfying $\n|_{\partial B}=\hat\n$ that are partially regular (that is, smooth outside a closed subset of $B$ of Hausdorff dimension less than 1), while if $K_1\leq\min(K_2,K_3)$ there are infinitely many such partially regular solutions having non-negative second variation.

One can similarly study when pure twists of the form \eqref{puretwist} are minimizers. As in Section \ref{lav} consider the region $\om=(0,l_1)\times(0,l_2)\times (0,d)$ between two parallel plates given by $x_3=0$ and $x_3=d>0$,  with planar boundary conditions on the plates
\be 
\label{planarbcs}
(\n\otimes \n)(x_1,x_2,0)= \n_0\otimes\n_0, \;\;(\n\otimes\n)(x_1,x_2,d)=\n_d\otimes\n_d,
\ee
where $\n_0=(\cos\alpha,\sin\alpha,0),\;\n_d=(\cos\beta,\sin\beta,0)$ and $\alpha,\beta\in[0,\pi)$, and periodic boundary conditions
\be 
&&\hspace{-.3in}(\n\otimes\n)(0,x_2,x_3)=(\n\otimes\n)(l_1,x_2,x_3),\;(\n\otimes\n)(x_1,0,x_3)=(\n\otimes\n)(x_1,l_2,x_3)\nonumber\\ &&  \label{perbcs}
\ee
on the other faces, where we have expressed the boundary conditions in terms of $\n\otimes\n$ because $\pm\n$ are physically indistinguishable (for simplicity we didn't do this in Section \ref{lav}). Then we have 
\begin{thm}
\label{planarthm}
The minimum of $I(\n)$ among $\n\in H^1(\om;S^2)$ satisfying the boundary conditions \eqref{planarbcs}, \eqref{perbcs} is attained. If $K_2\leq \min(K_1,K_3)$ then any minimizer is a pure twist of the form
\be 
\label{twistform}
\n^*(\x)=\pm(\cos\theta(x_3), \sin\theta(x_3),0),
\ee 
where  
\begin{equation}
\label{solns1}
\theta(x_3)=\left\{\begin{array}{cc}\alpha + (\beta-\alpha)\frac{x_3}{d}&\mbox{ if } |\beta-\alpha|\leq\frac{\pi}{2}\\
\alpha+ (\beta-\alpha-\pi\,{\rm sign}\,(\beta-\alpha))\frac{x_3}{d} &\mbox{ if } |\beta-\alpha|\geq\frac{\pi}{2}
\end{array}\right. .
\end{equation}
\end{thm}
\noindent Thus up to a physically meaningless change of sign, there is a unique minimizer if $\n_0\cdot\n_d\neq 0$, and two minimizers if $\n_0\cdot\n_d=0$ (corresponding to twists in opposite directions). The proof, which will appear elsewhere, is a consequence of the identity \eqref{identity} and the fact that for $\n$ satisfying the boundary conditions
$$\int_\om(\tr (\nabla\n)^2-(\Div\n)^2)\,d\x=0.$$
The conclusion of the theorem is intuitively reasonable, since the condition $K_2\leq\min(K_1,K_3)$ says that it is energetically easier to twist than to splay or bend.

If $K_2>\min(K_1,K_3)$ then pure twists are not in general minimizers. For example, if $K_1<K_2=K_3$ a twist-bend equilibrium solution depending only on $x_3$ has less energy than any pure twist (see Leslie \cite{leslie1975}, Stewart \cite{stewart04}). Using a deflation (see \cite{farrell2014}) numerical scheme Adler {\it et al.} \cite{adleretal16} compute a variety of equilibrium solutions for some different cases, studying for example the bifurcation to twist-bend solutions as $K_2$ is increased for fixed $K_1,K_3$.  It would be interesting to understand why (the presumably infinitely many) less regular equilibria similar to those whose existence is proved  by Rivi\`ere and Hong are not found by the implementation of the deflation method.

\subsection{Landau - de Gennes}
Consider the problem of minimizing the free energy \eqref{Qfree} with the singular bulk potential
$$I(\Q)=\int_\Omega \left(\psi_B^s(\Q,\theta)+\psi_E(\Q,\nabla\Q)\right)\,d\x,$$
subject to $\Q|_{\partial\Omega}=\bar\Q$, where $\bar\Q$ is smooth and
$\lambda_{\rm min}(\bar\Q(\x))>-\frac{1}{3}$ for all $\x\in\partial\Omega$.

As explained at the end of Section \ref{singbulk}, if the inequalities \eqref{newLcondns} hold, then the minimum of $I(\Q)$ is attained. Let $\Q^*$ be a minimizer, so that in particular we have that  
\be 
\label{EL*}I(\Q^*)= \int_\Omega \left(\psi_B^s(\Q^*,\theta)+\psi_E(\Q^*,\nabla\Q)^*\right)\,d\x<\infty.
\ee
 Is $\Q^*$ smooth? This would be straightforward to 
prove if we knew that 
\be 
\label{baway}
\lambda_{\rm min}(\Q^*(\x))\geq -\frac{1}{3}+\varepsilon, \mbox{ for all }\x\in\Omega
\ee
for some constant $\varepsilon>0$. For then we could show that $\Q^*$ is a weak solution of the corresponding Euler-Lagrange equations and use elliptic regularity theory in the same way as Davis \& Gartland \cite{gartlanddavis}.

 But surely \eqref{baway} must be true, because why would it be good   for
the integrand in \eqref{EL*} to  be infinite somewhere, when $\Q^*$ a minimizer? 
However in fact this phenomenon often arises in the calculus of variations, as we have already seen for the hedgehog defect, for the model of cavitation discussed in Section \ref{lavrentiev}, and more surprisingly in the one-dimensional example 
mentioned there,
so it is a delicate matter to show that it cannot happen.

A proof of \eqref{baway} in the one-constant approximation $\psi_E(\Q,\nabla\Q)=L_1|\nabla\Q|^2$ is given in \cite{u9,j59}, but the method seems not to work for more general elastic constants, for which the question of whether \eqref{baway} holds remains open. However, without proving \eqref{baway}, Evans, Kneuss \& Tran \cite{EvansKneussTran2016} show as a consequence of a more general partial regularity result that $\Q^*$ is smooth outside a closed subset $E\subset\om$ of measure zero.
 Bauman \& Phillips \cite{baumanphillips2016} study the regularity problem in 2D, in particular proving \eqref{baway}   when $L_2=L_3=0$ and $L_1'>0$, where $L_1'$ is defined in \eqref{newL3}.

\section{Omissions}
This paper does not pretend to be a comprehensive survey of   mathematical work on liquid crystals. Some   notable omissions concern:\\

\noindent 1. {\it Dynamics}:\\ 
Here one goal is to understand the qualitative properties of solutions to the Ericksen-Leslie equations, which were briefly mentioned in Section \ref{dynamics}. These equations consist of a momentum equation generalizing the Navier-Stokes equations for motion of a linear viscous fluid, coupled to an evolution equation for the director $\n$. This system of equations poses additional challenges to those already present for the Navier-Stokes equations, for which it is famously not known whether solutions to initial-boundary value problems in 3D are smooth or unique. A particular added difficulty for the Ericksen-Leslie equations is how to handle the unit vector constraint $|\n|=1$. Thus the first attempts  to prove existence of solutions relaxed this constraint by replacing $W(\n,\nabla\n)$ by $W(\n,\nabla\n)+\ep^{-2}(|\n|^2-1)^2$ in the hope of recovering the constraint in the limit $\ep\to 0+$, and global  existence and partial regularity of weak solutions to a simplified version of the corresponding system of equations in 3D was proved in \cite{lin89a,linliu95,linliu96}.  

For  versions of the Ericksen-Leslie equations in 2D with the unit vector constraint there is a relatively complete global existence, uniqueness and regularity theory   (see, for example,  \cite{hong2011,hongxin2012,huanglinwang2014,linlinwang2010,linwang2010}). These papers ignore the moment of inertia of molecules; a short-time existence theory when this is included but without dissipation in the director equation is provided in \cite{chechkinetal2016b}.   For special initial data global existence of a weak solution in 3D to a simplified Ericksen-Leslie system was proved in  \cite{linwang2016}, and for the same system the existence of solutions developing a singularity in finite time was proved in \cite{huanglinliuwang2016}.  These papers mostly concern  the one-constant approximation. For a discussion of numerical approximation of solutions to the Ericksen-Leslie equations see  \cite{walkington2011}.

There are corresponding sets of dynamical equations when the order paranmeter is the $\Q$ tensor, such as the Beris-Edwards model \cite{berisedwards1994}. For recent work on the existence and properties of solutions see for example \cite{abelsdolzmannliu2014,abelsdolzmannliu2016,paicuzarnescu2011, paicuzarnescu2012}, and in the context of the singular bulk potential \cite{feireisletal2015,wilkinson2015}. For results on how to relate these models to the Ericksen-Leslie equations see \cite{wangzhangzhang2015}. 

The dynamical behaviour of liquid crystals has many complex aspects (see e.g. \cite{forestetal2007}) worthy of mathematical treatment.\\

\noindent 2. {\it Smectics.} The results highlighted in this paper mainly concern nematics. Among mathematical studies of other liquid crystal phases, smectics (briefly touched on in Sections \ref{lav}, \ref{molmodels}) have generated considerable recent interest. See, for example, \cite{almog2008,baumanphillipspark2015,biscarietal2007,calderer08continuum,chengphillips2015,Climent-Ezquerraetal2010,Climent-Ezquerraetal2012,ColbertKellyPhillips2013,garciaetal2016,garciajoo2012,garciajoo2014,hanetal,joophillips2007,kamiensantangelo2006,meizhang2015,pan2008,pan2014,parkcalderer2006,santangelo05curvature,segattiwu2011}.\\

\noindent 3. {\it Liquid crystal elastomers.} Liquid crystal elastomers are materials formed from polymers to which  are attached liquid crystal mesogens. In the case of  nematic elastomers the model of Bladon, Warner \& Terentjev \cite{bladonetal,warnerterentjev} is based on a free-energy density $\psi(\nabla\y,\n)$ depending  on the deformation gradient $\nabla\y$ and  $\n$, but not on $\nabla\n$ (so that Frank elasticity is ignored). Minimizing with respect to $\n$ leads to a free-energy density $W(\nabla\y)=\min_{|\n|=1}\psi(\nabla\y,\n)$ which was shown by De Simone \& Dolzmann \cite{desimonedolzmann}   not to be quasiconvex, leading to nonattainment of energy minimizers and an explanation of observed laminated structures \cite{kundlerfinkelmann} similar to those seen in martensitic phase transformations. For a selection of recent mathematical work in this area see \cite{agostinianidalmasodesimone2015,caldereretal2015,
cesanadesimone2009,cesanaplucinskybhattacharya2015}.\\

\noindent 4. {\it Topological aspects.} 
Liquid crystals traditionally have been an area in which geometry and mechanics meet topology, for example in the topological description of defects.
An interesting area of current activity in which topology plays a role is that of nematic shells (see, for example, \cite{ canevariramaswamymajumdar2016,canevarisegattiveneroni2015,rossovirgakralj2012,segattietal2014,segattietal2016}). Related to this work is the interest in the topology of disclination lines induced by colloidal systems (see, for example, \cite{AlamaBronsardLamy2016,coparetal2015,martinezetal2014,tasinkevychetal2014,tkalecetal2011}).  Other interesting connections with topology appear in \cite{ackermansmalyukh2016a,ackermansmalyukh2016}. 
    
 \section*{Acknowledgements} This research  was supported by 
EPSRC
(GRlJ03466, the Science and Innovation award to the Oxford Centre for Nonlinear
PDE EP/E035027/1, and EP/J014494/1), the European Research Council under the European Union's Seventh Framework Programme
(FP7/2007-2013) / ERC grant agreement no 291053 and
 by a Royal Society Wolfson Research Merit Award. 

I acknowledge with thanks the hospitality of the Liquid Crystal Research Institute at Kent State University, where much of this lecture was prepared.   I am grateful to Giacomo Canevari, Patrick Farrell, David Kinderlehrer, Fanghua Lin, Tom Lubensky, Peter Palffy-Muhoray, Dan Phillips, Michaela Vollmer and Arghir Zarnescu  for illuminating discussions and references. I especially thank Epifanio Virga and the referee for their careful reading of the article and  very useful suggestions and comments.
 
\bibliography{gen2,balljourn,ballconfproc,ballprep}
\bibliographystyle{abbrv}

\end{document}